\documentclass{article}
\usepackage{amsmath}
\usepackage{amsfonts}

\usepackage{amsmath,amssymb}
\usepackage{graphicx}
\usepackage{color}

\newtheorem{theorem}{Theorem}
\newtheorem{lemma}{Lemma}
\newtheorem{corollary}{Corollary}

\newtheorem{definition}{Definition}

\newtheorem{remark}{Remark}

\newtheorem{example}{Example}

\makeatletter

\newcommand{\Rmnum}[1]{\expandafter\@slowromancap\romannumeral #1@}
\makeatother

\newcommand{\ls}[1]
    {\dimen0=\fontdimen6\the\font\lineskip=#1\dimen0
     \advance\lineskip.5\fontdimen5\the\font
     \advance\lineskip-\dimen0
     \lineskiplimit=0.9\lineskip
     \baselineskip=\lineskip
     \advance\baselineskip\dimen0
     \normallineskip\lineskip\normallineskiplimit\lineskiplimit
     \normalbaselineskip\baselineskip
     \ignorespaces}

\begin{document}

\bibliographystyle{abbrv}

\title{The cross-correlation distribution of a $p$-ary $m$-sequence of period $p^{2m}-1$ and its decimation by $\frac{(p^{m}+1)^{2}}{2(p^{e}+1)}$}
\author{Yuhua Sun$^{1}$, Hui Li$^{1}$, Zilong Wang$^{1}$\\
$^1$ State Key Laboratory of Integrated Service Networks,\\
Xidian University,
Xi'an, 710071, Shaanxi,
China\\
Email: sunyuhua-1@163.com\ \ \ \ lihui@mail.xidian.edu.cn\\
wzlmath@gmail.com\\
}

 \maketitle


\thispagestyle{plain} \setcounter{page}{1}

\begin{abstract}
Let $n=2m$, $m$ odd, $e|m$, and $p$ odd prime with $p\equiv1\ \mathrm{mod}\ 4$. Let $d=\frac{(p^{m}+1)^{2}}{2(p^{e}+1)}$. In this paper, we study the cross-correlation between a $p$-ary $m$-sequence $\{s_{t}\}$ of period $p^{2m}-1$ and its decimation $\{s_{dt}\}$. Our result shows that the cross-correlation function is six-valued and that it takes the values in $\{-1,\ \pm p^{m}-1,\ \frac{1\pm p^{\frac{e}{2}}}{2}p^{m}-1,\ \frac{(1- p^{e})}{2}p^{m}-1\}$. Also, the distribution of the cross-correlation is completely determined.

\end{abstract}

\ls{1.5}
\section{Introduction}

One problem of considerable interest has been to find a decimation value $d$ such that the cross-correlation between a $p$-ary $m$-sequence $\{s_{t}\}$ of period $p^{n}-1$ and its decimation $\{s_{dt}\}$ is low. When $\mathrm{gcd}(d,p^{n}-1)=1$, the decimated sequence $\{s_{dt}\}$ is also an $m$-sequence of period $p^{n}-1$. Basic results on the cross-correlation between two $m$-sequences can be found in [1-3].
When $\mathrm{gcd}(d,p^{n}-1)\neq1$, the sequence $\{s_{dt}\}$ has period $\frac{p^{n}-1}{\mathrm{gcd}(d,p^{n}-1)}$. For this case, there also are many good results which can be found in [4-7].

In \cite{7}, for an odd prime $p$, Choi, Lim, No, and Chung  investigated into the cross-correlation of a $p$-ary $m$-sequence of period $p^{n}-1$ and its decimated sequence by $d=\frac{(p^{m}+1)^{2}}{2(p+1)}$, where $n=2m$ and $m$ is odd. They have shown that the magnitude of the cross-correlation values is upper bounded by $\frac{p+1}{2}p^{\frac{n}{2}}+1$.
Recently, for $n=2m$, $p\equiv3\ \mathrm{mod}\ 4$, Luo, Helleseth, and Kholosha \cite{8} determined the distribution of the cross-correlation values of a $p$-ary $m$-sequence $\{s_{t}\}$ of period $p^{n}-1$ and its decimated sequence $\{s_{dt}\}$ by $d=\frac{(p^{m}+1)^{2}}{2(p^{e}+1)}$, where $m$ is odd and $e|m$. They derived that the cross-correlation is six-valued.

It is not difficult to observe that the decimation value $d$ in \cite{8} is a generalization of that in \cite{7}, i.e., for $p\equiv3\ \mathrm{mod}\ 4$ the result in \cite{7} can be generalized to a more general form. In fact, it can also  be generalized to the case of $p\equiv1\ \mathrm{mod}\ 4$. In this paper, for $p\equiv1\ \mathrm{mod}\ 4$, we will combine the machinery in \cite{10},\cite{8} and \cite{7} to study the cross-correlation between a $p$-ary $m$-sequence $\{s_{t}\}$ of period $p^{2m}-1$ and its decimated sequence $\{s_{dt}\}$ by the same $d$ as that in \cite{8}, that is, $d=\frac{(p^{m}+1)^{2}}{2(p^{e}+1)}$, $e|m$ and $m$ is odd. We show that the cross-correlation is also six-valued. And the distribution of the cross correlation values is determined.

\section{Preliminaries}
For an odd prime $p$,
let $\mathrm{F}_{p^{n}}$ denote the finite field with $p^{n}$ elements and $\mathrm{F}_{p^{n}}^{*}=\mathrm{F}_{p^{n}}\backslash\{0\}$.
The trace function $\mathrm{Tr}_{m}^{n}$ from the field $\mathrm{F}_{p^{n}}$ onto the subfield $\mathrm{F}_{p^{m}}$ is defined by
$$\mathrm{Tr}_{m}^{n}(x)=x+x^{p^{m}}+x^{p^{2m}}+\cdots+x^{p^{(h-1)m}},$$
where $h=n/m$.

We will use the following notations in the remaining part of this paper unless otherwise specified.
Let $p\equiv1\ \mathrm{mod}\ 4$, $n=2m$, $e|m$ and $d=\frac{(p^{m}+1)^{2}}{2(p^{e}+1)}$, where $m$ is an odd integer. Let $\alpha$ be a primitive element of $\mathrm{F}_{p^{n}}$.  Then we have $\mathrm{gcd}(d,p^{n}-1)=\frac{p^{m}+1}{2}$ and $d(p^{m+e}+1)\equiv p^{m}+1\ (\mathrm{mod}\ p^{n}-1)$ (These also can refer to \cite{7} or \cite{8}). It should be pointed out that both $\frac{p^{m}+1}{2}$ and $d$ are odd when $p\equiv1\ \mathrm{mod}\ 4$.

 A $p$-ary $m$-sequence $\{s_{t}\}$ is given by
$$s_{t}=\mathrm{Tr}_{1}^{n}(\alpha^{t}),$$
where $\mathrm{Tr}_{1}^{n}$ is the trace function from $\mathrm{F}_{p^{n}}$ onto $\mathrm{F}_{p}$.

 The periodic cross correlation function $C_{d}(\tau)$ between$ \{s_{t}\}$ and $\{s_{dt}\}$ is defined by
$$C_{d}(\tau)=\sum\limits_{t=0}^{p^{n}-2}\omega^{s_{t+\tau}-s_{dt}},$$
where $0\leq \tau \leq p^{n}-2$ and $\omega$ is a primitive complex $p$th root of unity.
\begin{definition}(\cite{9})
A quadratic form $x_{1},x_{2},\ldots,x_{n}$ in $n$ indeterminates over $\mathrm{F}_{p}$ is a homogeneous polynomial in $F_{p}[x_{1},x_{2},\ldots,x_{n}]$ of degree 2, that is,
$$f(x_{1},x_{2},\ldots,x_{n})=\sum\limits_{i,j=1}^{n}a_{i,j}x_{i}x_{j}\ \ \mathrm{with}\ a_{i,j}=a_{j,i}\in \mathrm{F}_{p}.$$
\end{definition}
The $n\times n$ matrix $A$ whose $(i,j)$ entry is $a_{i,j}$ is called the coefficient matrix of $f$. From \cite{9}, we know that every quadratic form over $\mathrm{F}_{p}$ is equivalent to a diagonal quadratic form $a_{1}x_{1}^{2}+a_{2}x_{2}^{2}+\cdots+a_{r}x_{r}^{2}$ over $F_{p}$, where $r\leq n$ is called the rank of $f$. For any $b\in F_{p}$, the number of solutions of $f(x_{1},x_{2},\ldots,x_{n})=b$ is equal to the number of solutions of $a_{1}x_{1}^{2}+a_{2}x_{2}^{2}+\cdots+a_{r}x_{r}^{2}=b$. We denote $\triangle=a_{1}a_{2}\cdots a_{r}$, then $\triangle$ is called the determinant of $f$.
\begin{definition}(\cite{9})
The quadratic character of $\mathrm{F}_{p^{n}}$ is defined as
$$ \eta(x)=\left\{
\begin{array}{lll}
\ 1,\ \ \ \mathrm{if}\  x \mathrm{\ is\  a \ nonzero\  square\  in}\  \mathrm{GF}(p^{n})\\
-1,\ \ \mathrm{if}\  x \ \mathrm{is\  a\  nonsquare\  in}\  \mathrm{GF}(p^{n})\\
\ 0,\ \ \ \ \mathrm{if}\  x=0.
\end{array}
\right. $$
\end{definition}
\begin{definition}(\cite{9})
The canonical additive of $\mathrm{F}_{p^{n}}$ is defined as
$$\chi(x)=\sum\limits_{x\in \mathrm{F}_{p^{n}}}\omega^{Tr_{1}^{n}(x)}.$$
\end{definition}
\begin{definition}(\cite{9})
The Gaussian sum $G(\eta,\chi)$ of $\eta$ and $\chi$ is defined as
$$G(\eta,\chi)=\sum\limits_{x\in F_{p^{n}}^{\ast}}\eta(x)\chi(x).$$
\end{definition}

\section{The ranks of two quadratic forms}

In this section, we will give some results to prove our main theorem. First, using the similar method of \cite{7} or \cite{8}, we can get the following four lemmas.

\begin{lemma}
Let the symbols be defined as in section 2. Then the cross-correlation between $ \{s_{t}\}$ and $\{s_{dt}\}$ is given by
\begin{align}
C_{d}(\tau)=-1+C(-1,c)&=-1+\sum\limits_{x\in F_{p^{n}}}\chi(-x^{d}+cx)\nonumber\\
&=-1+\frac{1}{2}\left(E(-1,c)+E(-\alpha^{d},c\alpha)\right)\nonumber
\end{align}
where $c=\alpha^{\tau}$, and
$
E(a,b)=\sum\limits_{x\in \mathrm{F}_{p^{n}}}\chi(ax^{p^{m}+1}+bx^{p^{m+e}+1}).
$
Further, $q_{a,b}(x)=\mathrm{Tr}_{1}^{n}(ax^{p^{m}+1}+bx^{p^{m+e}+1})$
is a quadratic form over $\mathrm{F}_{p}$.\ \ \ \ \ \ \ \ \ \ \ \ \ \ \ \ \ \ \ \ \ \ \ \ \ \ \ \ \ \ \ \ \ \ \ \ \ \ \ \ \ \ \ \ \ \ \ \ \ \ \ \ \ \ \ \ \ \ \ \ \ \ \ \ \ \ \ \ \ \ \ \ \ \ \ \ \ \ \ \ \ \ \ \ \ \ \ \ \ \ \ $\Box$
\end{lemma}

Let $r_{a,b}$ be the rank of the quadratic form $q_{a,b}(x)$. Then the following Lemma can be derived from Corollary 5 of \cite{7}.

\begin{lemma}
Let $p\equiv1\ \mathrm{mod}\ 4$. Let $\triangle$ is the determinant of $q_{a,b}(x)$, where $x\in \mathrm{F}_{p^{n}}$.  Then we have
$$E(a,b)=\sum\limits_{x\in \mathrm{F}_{p}^{n}}\omega^{q_{a,b}(x)}=\pm p^{n-\frac{r_{a,b}}{2}},$$
where $\pm$ depends on $\eta(\triangle)$ and $\eta$ is the quadratic character of $F_{p}$.\ \ \ \ \ \ \ \ \ \ \ \ \ \ \ \ \ \ \ \ \ \ \ \ \ \ \ \ \ \ \ \ \ \ \ \ \ \ \ \ $\Box$
\end{lemma}

By Lemmas 1 and 2, in order to compute $C_{d}(\tau)$, we need to find the values of $r_{-1,c}$ and $r_{-\alpha^{d},c\alpha}$ . The following Lemma which can be found in \cite{7} or \cite{4} gives us a method to compute them.

\begin{lemma}(\cite{7},\cite{4})
Let $f(x)\in \mathrm{F}_{p^{n}}[x]$ can be expressed as a quadratic form in $\mathrm{GF}(p)[x_1, x_2, \cdots x_n]$, where $x\in \mathrm{F}_{p^{n}}$. Furthermore, let
$$Y=\{y\in\mathrm{F}_{p^{n}}:f(x+y)=f(x)\  \mathrm{for}\  \mathrm{all}\  x\in \mathrm{F}_{p^{n}}\}.$$
Then rank($f$)=$n-\mathrm{Log}_{p}|Y|$.\ \ \ \ \ \ \ \ \ \ \ \ \ \ \ \ \ \ \ \ \ \ \ \ \ \ \ \ \ \ \ \ \ \ \ \ \ \ \ \ \ \ \ \ \ \ \ \ \ \ \ \ \ \ \ \ \ \ \ \ \ \ \ \ \ \ \ \ \ \ \ \ \ \ \ \ \ \ \ \ \ \ \ \ \ \ \ \ \ \ \ $\Box$
\end{lemma}

Using Lemma 3 and the method of \cite{7} or \cite{8}, we can get the following lemma.

\begin{lemma}
Let the symbols be defined as before. Then the number of solutions of $q_{a,b}(x+y)=q_{a,b}(x)$ for all $x\in F_{p^{n}}$ is equal to the number of solutions of
\begin{align}
b^{p^{m+e}}y^{p^{2e}}+\left(a^{p^{m+e}}+a^{p^{e}}\right)y^{p^{e}}+by=0 \label{1}
\end{align}
in $\mathrm{F}_{p^{n}}$. Furthermore, Since Eq.(\ref{1}) is a $\mathrm{F}_{p^{e}}$-linearized polynomial, we know that the number of the solutions of it is $1$, $p^{e}$ or $p^{2e}$. Hence, $r_{a,b}=n$, $n-e$ or $n-2e$.\ \ \ \ \ \ \ \ \ \ \ \ \ \ \ \ \ \ \ \ \ \ \ \  \ \ \ \ \ \ \ \ \ \ \ \ \ \ \ \ \ \ \ \ \ \ \ \ \ \ \ \ \ \ $\Box$
\end{lemma}

In fact, we can get further result about $E(-\alpha^{d},c\alpha)$. To this end, we need the following lemmas and corollary.

\begin{lemma}(Theorem 5.30 of \cite{9})
Let the symbols be defined as in section 2. Then
$$\sum\limits_{x\in F_{p^{n}}}\chi(ax^{2})=\eta(a)G(\eta,\chi),$$
where $a\in \mathrm{F}^{\ast}$.\ \ \ \ \ \ \ \ \ \ \ \ \ \ \ \ \ \ \ \ \ \ \ \  \ \ \ \ \ \ \ \ \ \ \ \ \ \ \ \ \ \ \ \ \ \ \ \ \ \ \ \ \ \ \ \ \ \ \ \ \ \  \ \ \ \ \ \ \ \ \ \ \ \ \ \ \ \ \ \ \ \ \ \ \ \ \ \ \ \ \ \ \ \ \ \ \ \ \ \  \ \ \ \ \ \ \ \ \ \ \ \ \ \ \ \ \ \ \ \ \ \ \ \ \ \ \ \ \ \ $\Box$
\end{lemma}

\begin{lemma}(Theorem 5.16 of \cite{9})
Let the symbols be defined as in section 2. Then
$$
G(\eta,\chi)=\left\{
\begin{array}{ll}
\ p^{m}\ \ \ \ \mathrm{if}\ \frac{p^{m}+1}{2}\ \mathrm{even},\\
-p^{m}\ \ \ \mathrm{if}\ \frac{p^{m}+1}{2}\ \mathrm{odd}.
\end{array}
\right.
$$
\ \ \ \ \ \ \ \ \ \ \ \ \ \ \ \ \ \ \ \ \ \ \ \  \ \ \ \ \ \ \ \ \ \ \ \ \ \ \ \ \ \ \ \ \ \ \ \ \ \ \ \ \ \ \ \ \ \ \ \ \ \  \ \ \ \ \ \ \ \ \ \ \ \ \ \ \ \ \ \ \ \ \ \ \ \ \ \ \ \ \ \ \ \ \ \ \ \ \ \  \ \ \ \ \ \ \ \ \ \ \ \ \ \ \ \ \ \ \ \ \ \ \ \ \ \ \ \ \ \ $\Box$
\end{lemma}

\begin{corollary}
Let the symbols be defined as in section 2. Then we have
$$\sum\limits_{x\in F_{p^{n}}}\chi(a x^{2})=-\eta(a)p^{m},$$
where $a\in \mathrm{F}^{\ast}$.
\end{corollary}
{\bf Proof:} Note that $\frac{p^{m}+1}{2}$ is odd when $p\equiv1\ \mathrm{mod}\ 4$. Combining Lemmas 5 and 6, the result follows. \ \ \ \ $\Box$

\begin{lemma}
Let the symbols be defined as before. Then\\
(1) $E(-\alpha^{d},c\alpha)=\eta(c)p^{m}$.\\
(2) $r_{-\alpha^{d},c\alpha}=n$.
\end{lemma}
{\bf Proof:} (1) Similar to the proof of Lemma 1 in \cite{8}, we can get
$$E(-\alpha^{d},c\alpha)=\sum\limits_{x\in F_{p^{n}}}\chi(c\alpha x^{2}).$$
Note that $\alpha$ is a nonsquare in $\mathrm{F}_{p^{n}}$. By Corollary 1, we can get
$$E(-\alpha^{d},c\alpha)=-\eta(c\alpha)p^{m}=(-1)\eta(\alpha)\eta(c)p^{m}=\eta(c)p^{m}.$$
(2) By lemma 2 and the above result, we can get that $r_{-\alpha^{d},c\alpha}=n$.\ \ \ \ \ \ \ \ \ \ \ \ \ \ \ \ \ \ \ \ $\Box$

Combining Lemmas 1, 2, 4 and 7, we can get the following corollary.

\begin{corollary}
Let the symbols be defined as above. Then we have
$$E(-1,c)\in \{\pm p^{m},\ \pm p^{m+\frac{e}{2}},\ \pm p^{m+e}\},\ E(-\alpha^{d},c\alpha)\in\{\pm p^{m}\}$$
and
$$C(-1,c)\in \{0,\ \pm p^{m},\ \frac{1\pm p^{\frac{e}{2}}}{2}p^{m},\ \frac{-1\pm p^{\frac{e}{2}}}{2}p^{m},\ \frac{\pm(1- p^{e})}{2}p^{m},\ \frac{\pm(1 +p^{e})}{2}p^{m}\}.$$
\ \ \ \ \ \ \ \ \ \ \ \ \ \ \ \ \ \ \ \ \ \ \ \ \  \ \ \ \ \ \ \ \ \ \ \ \ \ \ \ \ \ \ \ \ \ \ \ \ \ \ \ \ \ \ \ \ \ \ \ \ \ \ \ \ \ \ \ \ \ \ \ \ \ \ \ \ \ \ \ \ \ \ \ \ \ \ \ \ \ \ \ \ \ \ \ \ \ \ \ \ \ \ \ \ \ \ \ \ \ \ \ \ \ \ \ \ \ \ \ \ \ \ $\Box$
\end{corollary}

In fact, we can prove $C(-1,c)\neq\frac{-1\pm p^{\frac{e}{2}}}{2}p^{m}$. Before proving it, we need the following lemma.
\begin{lemma}
( \cite{5} ) Let $g_{\upsilon}(z)=z^{p^{e}+1}-\upsilon z+\upsilon$, where $\upsilon\in F_{p^{n}}^{\ast}$. Then the following results hold.\\
(1) The Equation $g_{\upsilon}(z)=0$ has either 0, 1, 2, or $p^{e}+1$ roots in $F_{p^{n}}$.\\
(2) If $g_{\upsilon}(z)=0$ has only one root in $F_{p^{n}}$, then $c$ is a square element in $F_{p^{n}}$;\\
(3) Let $N_{1}$ denote the number of $\upsilon\in F_{p^{n}}^{\ast}$ such that $g_{\upsilon}(z)=0$ has only one root in $F_{p^{n}}$. Then
$N_{1}=p^{n-e}.$\\
(4) If $g_{\upsilon}(z)=0$ has only one root $z_{0}$ in $F_{p^{n}}$, then $z_{0}$ satisfies $(z_{0}-1)^{\frac{p^{n}-1}{p^{e}-1}}=1,$
i.e., $z_{0}-1$ is a $(p^{e}-1)$th power in $F_{p^{n}}.$\ \ \ \ \ \ \ \ \ \ \ \ \ \ \ \ \ \ \ \ \ \ \ \ \ \ \ \ \ \ \ \ \ \ \ \ \ \ \ \ \ \ \ \ \ \ \ \ \ \ \ \ \ \ \ \ \ \ \ \ \ \ \ \ \ \ \ \ \ \ \ \ \ \ \ \ \ \ \ \ \ \ \ \ \ \ \ \ \ \ \ \ \ \ \ \ \ \ \ \ \ \ \ \ \ \ \ \ \ \ \ \ \ \ \ $\Box$
\end{lemma}

\begin{lemma}
Let the symbols be defined as before. Then the following two results hold.\\
(1) If $r_{-1,c}=n-e$, then $c$ is a square in $F_{p^{n}}^{\ast}$ and $E(-\alpha^{d},c\alpha)=p^{m}$. Consequently, we get $C(-1,c)\neq\frac{-1\pm p^{\frac{e}{2}}}{2}p^{m}$.\\
(2) Let $N_{e}$ be the number of $c^{'}$s such that $r_{-1,c}=n-e$, where $c\in F_{p^{n}}^{\ast}$. Then $N_{e}=p^{m-e}(p^{m}+1)$.
\end{lemma}
{\bf Proof:} (1) Let $f_{c}(x)=c^{p^{m+e}}x^{p^{2e}}-2x^{p^{e}}+cx$, $y=x^{p^{e}-1}$ and $z=\frac{2}{c}y$. By Lemma 4, for $r_{-1,c}=n-e$, we know that the equation $f_{c}(x)=0$  has $p^{e}-1$ nonzero solutions in $F_{p^{n}}^{\ast}$ and that any two such solutions $x_{1}$, $x_{2}$ satisfy $x_{1}^{p^{e}-1}=x_{2}^{p^{e}-1}$, i.e.,
 $g_{c}(y)=c^{p^{m+e}}y^{p^{e}+1}-2y+c=0$ has only one solution $y_{0}$ and $y_{0}$ is a $(p^{e}-1)$th power in $F_{p^{n}}^{\ast}$. Further, $g_{c}(y)=0$ has only one solution in $F_{p^{n}}^{\ast}$ if and only if $g_{\upsilon}(z)=0$ has only one solution $z_{0}\in F_{p^{n}}^{\ast}$, where
 $\upsilon=\frac{4}{c^{(p^{m}+1)p^{e}}}$. Note that $c=\alpha^{\tau}$ and that $\beta=\alpha^{p^{m}+1}$ is a primitive element in $F_{p^{m}}$. Then we have $c^{p^{m}+1}=\beta^{\tau}$ and $\upsilon=4\beta^{-\tau p^{e}}\in F_{p^{m}}^{\ast}$. By the same argument as that of Lemma 4 in \cite{8}, we can get that $g_{\upsilon}(z)=0$ has only one root in $F_{p^{n}}^{\ast}$ if and only if $g_{\upsilon}(z)=0$ has only one root in $F_{p^{m}}^{\ast}$. By Lemma 8 (2), when $g_{\upsilon}(z)=0$ has only one root in $F_{p^{m}}^{\ast}$ the element $\upsilon$ is a square in $F_{p^{m}}^{\ast}$, which implies that $\tau$ is even and that $c$ is a square in $F_{p^{n}}$.

 (2) Since $\upsilon=\frac{4}{c^{(p^{m}+1)p^{e}}}=\left(\frac{2}{c^{p^{e}}}\right)^{p^{m}+1}\in F_{p^{m}}^{\ast}$, then $\upsilon$ runs through $p^{m}+1$ times all the elements in $F_{p^{m}}^{\ast}$ when $c$ runs through all the nonzero element in $F_{p^{2m}}^{\ast}$.  By Lemma 8 (3), there are $p^{m-e}$ $\upsilon^{'}$s such that $g_{\upsilon}(z)=0$ has only one root in $F_{p^{m}}^{\ast}$ when $\upsilon$ runs through all the elements in $F_{p^{m}}^{\ast}$. Hence, there are $p^{m-e}(p^{m}+1)$ $c^{'}$s in $F_{p^{2m}}^{\ast}$ such that $g_{\upsilon}(z)=0$ has only one root in $F_{p^{m}}^{\ast}$. By the argument of (1), there also are $p^{m-e}(p^{m}+1)$ $c^{'}$s such that $g_{c}(y)=0$ has only one root in $F_{p^{n}}^{\ast}$. Let $c_{0}$ be any one of $p^{m-e}(p^{m}+1)$ $c^{'}$s and let $y_{0}\in F_{p^{2m}}^{\ast}$ be the only root of $g_{c_{0}}(y)=0$. Next, we will prove that $y_{0}$ must be a $(p^{e}-1)$th power in $F_{p^{2m}}^{\ast}$, i.e., $y_{0}^{\frac{p^{2m}-1}{p^{e}-1}}=1$. Let $z_{0}=\frac{2}{c_{0}}y_{0}\in F_{p^{m}}^{\ast}$. Then $z_{0}^{p^{e}+1}-\frac{4}{c_{0}^{(p^{m}+1)p^{e}}}z_{0}+\frac{4}{c_{0}^{(p^{m}+1)p^{e}}}=0$, i.e., $z_{0}-1=\frac{c_{0}^{(p^{m}+1)p^{e}}z_{0}^{p^{e}+1}}{4}$. By Lemma 8 (4), $(z_{0}-1)^{\frac{p^{m}-1}{p^{e}-1}}=\left(\frac{c_{0}^{(p^{m}+1)p^{e}}z_{0}^{p^{e}+1}}{4}\right)^{\frac{p^{m}-1}{p^{e}-1}}=\left(\frac{c_{0}^{p^{m}+1}z_{0}^{2}}{4} \right)^{\frac{p^{m}-1}{p^{e}-1}}=\left(\frac{c_{0}^{p^{m}+1}(\frac{2}{c_{0}}y_{0})^{2}}{4} \right)^{\frac{p^{m}-1}{p^{e}-1}}=\left(c_{0}^{p^{m}-1}y_{0}^{2}\right)^{\frac{p^{m}-1}{p^{e}-1}}=1$, which implies that there exists an element $\theta\in \mathrm{F}_{p^{m}}$ such that $c_{0}^{p^{m}-1}y_{0}^{2}=\theta^{p^{e}-1}$, i.e., $y_{0}^{\frac{p^{n}-1}{p^{e}-1}}=\left[\frac{\theta^{p^{e}-1}}{c_{0}^{p^{m}-1}}\right]^{\frac{p^{n}-1}{2(p^{e}-1)}}=\frac{\theta^{(p^{m}-1)\cdot\frac{p^{m}+1}{2}}}{c_{0}^{\frac{p^{m}-1}{p^{e}-1} \frac{p^{n}-1}{2}}}=\frac{1}{\left[c_{0}^{\frac{p^{n}-1}{2}}\right]^{\frac{p^{m}-1}{p^{e}-1}}}=1$. By the result of (1), we know that $c_{0}$ is a square in $F_{p^{n}}$, then $c_{0}^{\frac{p^{n}-1}{2}}=1$. Hence, we have $y_{0}^{\frac{p^{n}-1}{p^{e}-1}}=1$, i.e., $y_{0}$ is a $(p^{e}-1)$th power in $F_{p^{n}}$. The result follows.\ \ \ \ \ \ \ \ \ \ \ \ \ \ \ \ \ \ \ \ \ \ \ \ \ \ \ \ \ \ \ \ \ \ \ \ \ \ \ \ \ \ \ \ \ \ \ \ \ \ \ \ \ \ \ \ \ \ \ \ \ \ \ \ \ \ \ \ \ \ \ \ \ \ \ \ \ \ \ \ \ \ \ \ $\Box$

In the next section, we will find the equations which are satisfied by $E(-1,c)$ and $C(-1,c)$ in order to determine the distribution of $C(-1,c)$ or $C_{d}(\tau)$.

\section{The distribution of $C_{d}(\tau)$}
First, using the method of \cite{10},we can get the following Lemma.

\begin{lemma}
Let the symbols be defined as above. We have the following results:\\
(1) $\sum\limits_{c\in F_{p^{2m}}}E(-1,c)=p^{2m}$; $E(-1,0)=-p^{m}$;\\
(2) $\sum\limits_{c\in F_{p^{2m}}}\left[E(-1,c)\right]^{2}=(2p^{2m}-1)p^{2m}$;\\
(3) $\sum\limits_{c\in F_{p^{2m}}}C(-1,c)=p^{2m}$; $C(-1,0)=\frac{1}{2}(p^{m}-1)p^{m}$;\\
(4) $\sum\limits_{c\in F_{p^{2m}}}\left[C(-1,c)\right]^{2}=p^{4m}$.\ \ \ \ \ \ \ \ \ \ \ \ \ \ \ \ \ \ \ \ \ \ \ \ \ \ \ \ \ \ \ \ \ \ \ \ \ \ \ \ \ \ \ \ \ \ \ \ \ \ \ \ \ \ \ \ \ \ \ \ \ \ \ \ \ \ \ \ \ \ \ \ \ \ \ \ \ \ \ \ \ \ \ \ $\Box$
\end{lemma}

\begin{corollary}
Let $N_{i}=|\{c\in \mathrm{F}_{p^{n}}^{\ast}|r_{-1,c}=n-i\}|$, where $i=0,e,2e$. Then we have
\begin{align}
&N_{0}=\frac{(p^{m+2e}-p^{m+e}-p^{m}-p^{2e}+2)(p^{m}+1)}{p^{2e}-1},\nonumber\\
&N_{e}=p^{m-e}(p^{m}+1),\nonumber\\
&N_{2e}=\frac{(p^{m-e}-1)(p^{m}+1)}{p^{2e}-1}.\nonumber
\end{align}
\end{corollary}
{\bf Proof:} By Lemma 9 (2), we know that $N_{e}=p^{m-e}(p^{m}+1)$. Note that $N_{0}+N_{e}+N_{2e}=p^{2m}-1.$ Further, by Lemma 10 (2), we have
$$N_{0}+N_{e}p^{e}+N_{2e}p^{2e}=2(p^{2m}-1).$$
Straightforward calculation gives the result.\ \ \ \ \ \ \ \ \ \ \ \ \ \ \ \ \ \ \ \ \ \ \ \ \ \ \ \ \ \ \ \ \ \ \ \ \ \ \ \ \ \ \ \ \ \ \ \ \ \ \ \ \ \ \ \ \ \ \ \ \ \ \ \ \ \ \ \ \ \ \ \ \ \ \ \ \ \ \ \ \ \ \ \ $\Box$

Combining the machinery in \cite{10}, \cite{8} and \cite{7}, we have the following results.

\begin{lemma}
Let the symbols be defined as above. Then\\
(1)$\sum\limits_{c\in F_{p^{2m}}}\left[E(-1,c)\right]^{3}=(-p^{2m}+p^{m+e}+p^{e})p^{3m}.$\\
(2)$\sum\limits_{c\in F_{p^{2m}}}\left[C(-1,c)\right]^{3}=\frac{1}{8}p^{3m}(p^{3m}-p^{2m}+p^{m+e}+6p^{m}+p^{e}).$
\end{lemma}

\begin{corollary}
Let
$N_{i,\epsilon}=|\{c\in \mathrm{F}_{p^{n}}^{\ast}|E(-1,c)=\epsilon p^{m+\frac{i}{2}}\}|,$
where $i=0,e,2e$ and $\epsilon=\pm1$. Then we have
\begin{align}
&N_{0,1}=\frac{(p^{m+e}-1)(p^{m}+1)}{2(p^{e}+1)},\ \ N_{0,-1}=\frac{(p^{m+e}-2p^{m}-2p^{e}+3)(p^{m}+1)}{2(p^{e}-1)},\nonumber\\
&N_{e,1}=N_{e,-1}=\frac{1}{2}p^{m-e}(p^{m}+1),\nonumber\\
&N_{2e,1}=0,\ N_{2e,-1}=\frac{(p^{m-e}-1)(p^{m}+1)}{p^{2e}-1}.\nonumber
\end{align}
\end{corollary}
{\bf Proof:} Using Corollary 3, Lemma 10 (1) and Lemma 11 (1),  we know that
\begin{align}
&N_{0,1}+N_{0,-1}=N_{0}=\frac{(p^{m+2e}-p^{m+e}-p^{m}-p^{2e}+2)(p^{m}+1)}{p^{2e}-1},\nonumber\\
&N_{e,1}+N_{e,-1}=N_{e}=p^{m-e}(p^{m}+1),\nonumber\\
&N_{2e,1}+N_{2e,-1}=N_{2e}=\frac{(p^{m-e}-1)(p^{m}+1)}{p^{2e}-1}.\nonumber\\
&(N_{0,1}-N_{0,-1})+p^{\frac{e}{2}}(N_{e,1}-N_{e,-1})+p^{e}(N_{2e,1}-N_{2e,-1})=p^{m}+1,\label{16}\\
&(N_{0,1}-N_{0,-1})+p^{\frac{3e}{2}}(N_{e,1}-N_{e,-1})+p^{3e}(N_{2e,1}-N_{2e,-1})=(p^{m}+1)(-p^{m}+p^{e}+1).\label{17}
\end{align}
Note that both $p^{\frac{e}{2}}$ and $p^{\frac{3e}{2}}$ are irrational numbers. Eqs. (\ref{16}) and (\ref{17}) imply that $N_{e,1}=N_{e,-1}$.
Straightforward calculation gives the result.\ \ \ \ \ \ \ \ \ \ \ \ \ \ \ \ \ \ \ \ \ \ \ \ \ \ \ \ \ \ \ \ \ \ \ \ \ \ \ \ \ \ \ \ \ \ \ \ \ \ \ \ \ \ \ \ \ \ \ \  \ \ \ \ \ \ \ \ \ \ \ \ \ \ \ \ \ \ \  \ \ \ \ \ \ \ \ \ \ \ \ \ \ \ \ \ \ \ $\Box$

\begin{theorem}
Let $n=2m$, $e|m$, where $m$ is odd. Let $p\equiv1\ \mathrm{mod}\ 4$ and $d=\frac{(p^{m}-1)^{2}}{2(p^{e}+1)}$. Then we get the distribution of the cross correlation $C_{d}(\tau)$ in the following.
\begin{align}
&-1\ \ \ \ \ \ \ \ \ \ \ \ \ \ \ \ \ \ \ \ \ \mathrm{occurs} \ \ \ \ \ \frac{(p^{m+e}-2p^{m}-2p^{e}+3)(p^{m}+1)}{2(p^{e}-1)}\ \ \mathrm{times}\nonumber\\
&p^{m}-1\ \ \ \ \ \ \ \ \ \  \ \ \ \ \ \ \ \ \mathrm{occurs}\ \ \ \ \ \frac{(p^{e}-1)(p^{m}+1)^{2}}{4(p^{e}+1)}\ \ \mathrm{times}\nonumber\\
&-p^{m}-1\ \ \ \ \ \ \ \ \ \ \ \ \ \ \mathrm{occurs} \ \ \ \ \ \ \frac{p^{2m}-1}{4}\ \ \mathrm{times}\nonumber\\
&\frac{1\pm p^{\frac{e}{2}}}{2}p^{m}-1\ \ \ \ \ \ \ \ \ \mathrm{occurs}\ \ \ \ \ \frac{1}{2}p^{m-e}(p^{m}+1)\ \ \mathrm{times}\nonumber\\
&\frac{1-p^{e}}{2}p^{m}-1\ \ \ \ \ \ \ \ \ \mathrm{occurs}\ \ \ \ \ \frac{(p^{m-e}-1)(p^{m}+1)}{p^{2e}-1}\ \ \mathrm{times}\nonumber
\end{align}
\end{theorem}
Proof: For convenience, we denote
\begin{align}
N_{i,\epsilon_{1},\epsilon_{2}}=|\{c\in F_{p^{2m}}^{\ast}|E(-1,c)=\epsilon_{1}p^{m+\frac{i}{2}}\ \mathrm{and}\ E(-\alpha^{d},c\alpha)=\epsilon_{2}p^{m}\}|,\nonumber
\end{align}
where $i\in\{0,e,2e\}$, $\epsilon_{1}=\pm1$, and $\epsilon_{2}=\pm1.$
By Lemma 9 (1), we know that
\begin{align}
N_{e,1,-1}=N_{e,-1,-1}=0.\label{42.1}
\end{align}
On the other hand, by Corollary 4, we know that
$$N_{e,1}=N_{e,1,1}+N_{e,1,-1}=N_{e,-1}=N_{e,-1,1}+N_{e,-1,-1}=\frac{1}{2}p^{m-e}(p^{m}+1),$$
then we can get
\begin{align}
N_{e,1,1}=N_{e,-1,1}=\frac{1}{2}p^{m-e}(p^{m}+1).\label{42.2}
\end{align}
Again, by Corollary 4, we can get
\begin{align}
N_{0,1,1}+N_{0,1,-1}&=N_{0,1}=\frac{(p^{m+e}-1)(p^{m}+1)}{2(p^{e}+1)},\label{43}\\
N_{0,-1,1}+N_{0,-1,-1}&=N_{0,-1}=\frac{(p^{m+e}-2p^{m}-2p^{e}+3)(p^{m}+1)}{2(p^{e}-1)},\label{44}\\
N_{2e,1,1}+N_{2e,1,-1}&=N_{2e,1}=0,\label{45}\\
N_{2e,-1,1}+N_{2e,-1,-1}&=N_{2e,-1}=\frac{(p^{m-e}-1)(p^{m}+1)}{p^{2e}-1}\label{46}
\end{align}
Further, by Lemma 10 (3),(4) and Lemma 11 (2), we can get
\begin{align}
(N_{0,1,1}-N_{0,-1,-1})+\frac{1-p^{e}}{2}N_{2e,-1,1}&-\frac{1+p^{e}}{2}N_{2e,-1,-1}\nonumber\\
&=\frac{1}{2}(-p^{m-e}+1)(p^{m}+1),\label{47}\\
(N_{0,1,1}+N_{0,-1,-1})+(\frac{1-p^{e}}{2})^{2}N_{2e,-1,1}&+(\frac{1+p^{e}}{2})^{2}N_{2e,-1,-1}\nonumber\\
&=\frac{1}{4}(2p^{m}-p^{m-e}-1)(p^{m}+1)\label{48}\\
(N_{0,1,1}-N_{0,-1,-1})+(\frac{1-p^{e}}{2})^{3}N_{2e,-1,1}&-(\frac{1+p^{e}}{2})^{3}N_{2e,-1,-1}\nonumber\\
&=\frac{1}{8}(2p^{m}+p^{e}+1)(p^{m}+1).\label{49}
\end{align}
Combining Eqs. (4.40-4.48), we can get the result.\ \ \ \ \ \ \ \ \ \ \ \ \ \ \ \ \ \ \ \ \ \ \ \ \ \ \ \ \ \ \ \ \ \ \ \ \ \ \ \ \
\ \ \ \ \ \ \ \ \ \ \ \ \ \ \ \ \ \ \ \ \ \ \ \ \ \ \ \ \ \ \ $\Box$

\begin{remark}
For $p=5$ and $e=1$ in Theorem 1, the magnitude of the cross-correlation is upper bounded by $2\sqrt{p^{n}}+1$. This is meaningful in CDMA communication systems.
\end{remark}

\end{document}